\documentstyle[11pt,newpasp,twoside,epsf]{article}

\def\spose#1{\hbox to 0pt{#1\hss}}
     
\begin{document}

\title{Brightest Cluster Galaxies and the Las Campanas Distant Cluster
Survey}

\author{Dennis Zaritsky}
\affil{Steward Observatory, 933 N. Cherry Ave., University of Arizona,
Tucson, AZ, 85721, USA}

\author{Amy E. Nelson}
\affil{Department of Astronomy and Astrophysics, Univ. of Calif. at Santa Cruz,
Santa Cruz, CA, 95064, USA}

\author{Luc Simard}
\affil{Steward Observatory, 933 N. Cherry Ave., University of Arizona,
Tucson, AZ, 85721, USA}

\author{Anthony H. Gonzalez}
\affil{Harvard-Smithsonian Center for Astrophysics, 60 Garden St., Cambridge, MA, 02138, USA}

\author{Julianne J. Dalcanton}
\affil{Department of Astronomy, University of Washington, Box 351580, Seattle, WA, 98195, USA}

\begin{abstract}
We present our study of the evolution of brightest cluster galaxies
based on data from the Las Campanas Distant Cluster Survey
(Gonzalez{\it et al.} 2001) and HST optical and IR imaging.
We briefly discuss the 
technique that enabled us to use short ($\sim 3$ min) exposures
and a small (1m) telescope to efficiently survey $\sim$ 130 sq. deg.
of sky for high-z galaxy clusters. 
Follow-up imaging is used to construct a sample of several
tens of BCGs at $0.3 \le z \le 0.9$ 
with which to explore their evolution. In particular,
we examine the luminosity and color evolution of these
galaxies. We confirm the previous results that 1) BCGs in 
different environments evolve differently (eg. Burke, Collins, 
\& Mann 2000), 2) BCGs, particularly those in low-mass clusters, 
must be accreting significantly since $z \sim 1$ (Aragon-Salamanca 
{\it et al.} 1998). From measurments of the colors vs. redshift,
we conclude that this accretion must consist of old stellar populations.
Using HST NICMOS and WFPC2 images, we 
find preliminary evidence for evolution in the sizes of BCGs
that is consistent with the accretion model presented in standard
hierarchical models (Aragon-Salamanca {\it et al.} 1998), but 
puzzling differences between NICMOS and WFPC2 scale lengths precludes
a definitive conclusion.
\end{abstract}

\keywords{Galaxy: clusters; Galaxy: evolution}

\section{Introduction}

While definition dictates that every galaxy cluster must have a brightest 
galaxy, these galaxies are not simply the bright tail of the luminosity
function (Oemler 1976; Tremaine \& Richstone 1977)). 
Locally, these galaxies are excellent standard candles
(Hoessel 1980) 
and they appear to remain so to $z \sim 1$ (Aragon-Salamanca et al.
1993; Collins, Burke, and Mann 2000). They appear to be intricately
tied to their host cluster because at least some have semimajor
axes that extend well beyond
500 kpc (cf. Gonzalez et al. 2000). The evolution of these galaxies
is therefore part galaxy evolution, part cluster evolution. The Las 
Campanas Distant Cluster Survey (LCDCS) provides a new, large sample
of clusters with which to construct significant samples of BCGs at
$z > 0.5$. 

\section{The Survey}

The basic premise of our cluster detection technique is that we utilize
the light from unresolved cluster galaxies. Therefore, we do not need
to obtain deep images of the sky to detect a statistically
significant number of cluster galaxies, rather we need an image that
contains a statistically significant number of cluster photons. The
method is described by Dalcanton (1996), Zaritsky {\it et al.} (1997),
and Gonzalez {\it et al.} (2001).

Our current survey is based on drift scan observations obtained with
the Las Campanas 1m Swope telescope and the Great Circle Camera
(Zaritsky, Bredthauer, \& Shectman 1996). We obtained two scans through
each region of the survey area, each with an effective exposure time
of $\sim$ 90 s. The power of the technique is evident in the detection
of clusters out to $z \sim 1.1$ with a 1m telescope and 3 min exposures.
Briefly, the technique involves several flat-fielding passes, 
masking of bright stars, removal of resolved galaxies and faint stars, 
and smoothing with a kernel that corresponds roughly to the size of
cluster cores at $z \sim 0.6$. 
We refer the interested reader to Gonzalez {\it et al.} (2001) for
a description of the data reduction techniques and a complete listing
of all the candidate clusters.

\section{Follow-Up Observations}

The most telescope-intensive part of the project has been the follow-up
observations aimed at testing our cluster candidates and calibrating
methods to estimate the cluster redshift and mass (see below). 
From imaging in $V, I$, and $K^\prime$, we construct a sample of
BCGs drawn from photometry of 63 confirmed galaxy clusters 
(not all are imaged in each filter).

With surveys that produce thousands of candidate galaxy
cluster identifications, it becomes impractical to obtain redshift
and mass measures via spectroscopy for a significant fraction of 
the candidates. Ideally, these quantities can
be estimated from the survey data themselves. In doing so, we will
necessarily
sacrifice precision on a cluster-by-cluster basis, but as long as 
the uncertainties are well understood, the sheer number of clusters 
allows high precision measures of the statistical properties of the sample. 
There really is no choice in this matter - 
to obtain a sufficient number of galaxy spectroscopic
redshifts for mass measurements of a sample of 
ONLY 20 clusters requires $\sim$ 35 nights on an 8 to 10m class
telescope.

We use cluster spectroscopic redshifts when available,
or photometric redshifts based on the color of the red sequence of
early-type galaxies in the cluster (Nelson {\it et al.} 2001a).
The calibration of the mass indicator is much more speculative
primarily because of the dearth of mass estimates for high redshift
clusters. Only one previously known X-ray cluster at $z > 0.35$ lies
within our survey area. To augment this, 
we did small drift scans around 17 other
clusters with X-ray luminosity and/or temperature measurements. 
Unfortunately, the relationships between $\Sigma$ and other mass
indicators ($L_x, T_x,$ or velocity dispersion) are somewhat 
poorly defined (Figure 1).
Two aspects are particularly vexing: 1) we need to determine not
only the relation between these quantities but we also need to 
quantify the scatter to successfully apply this method, and 2) most
of the data are for clusters at the low end of the redshift range.
Extracting the full potential of this, or any other
survey, is predicated on developing a reliable mass estimator and
understanding its uncertainties. Nevertheless, these correlations enable
us to split the sample into likely low-L$_X$ and high-L$_X$ subsamples
to investigate the role of environment on BCG properties. 

\begin{figure}
\plottwo{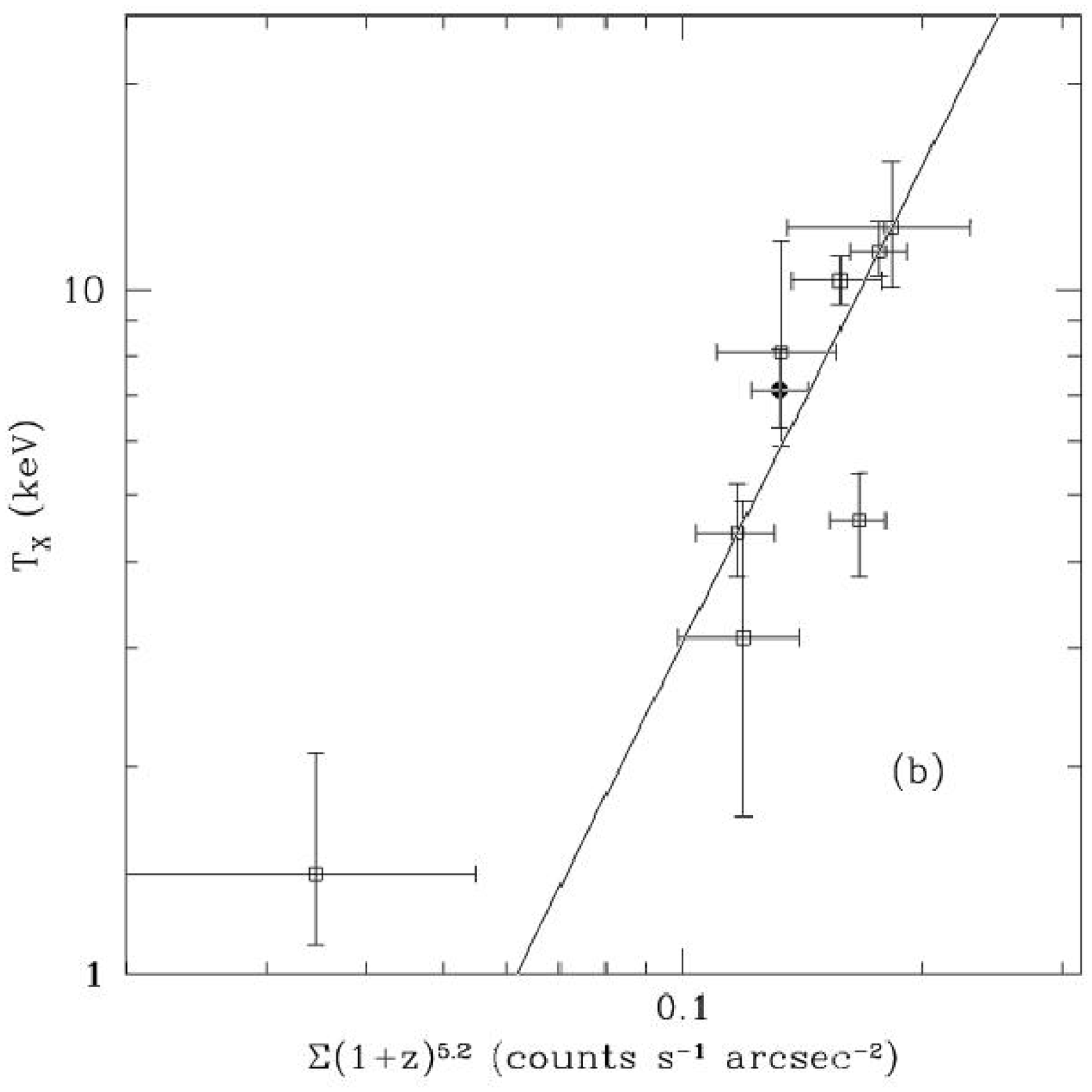}{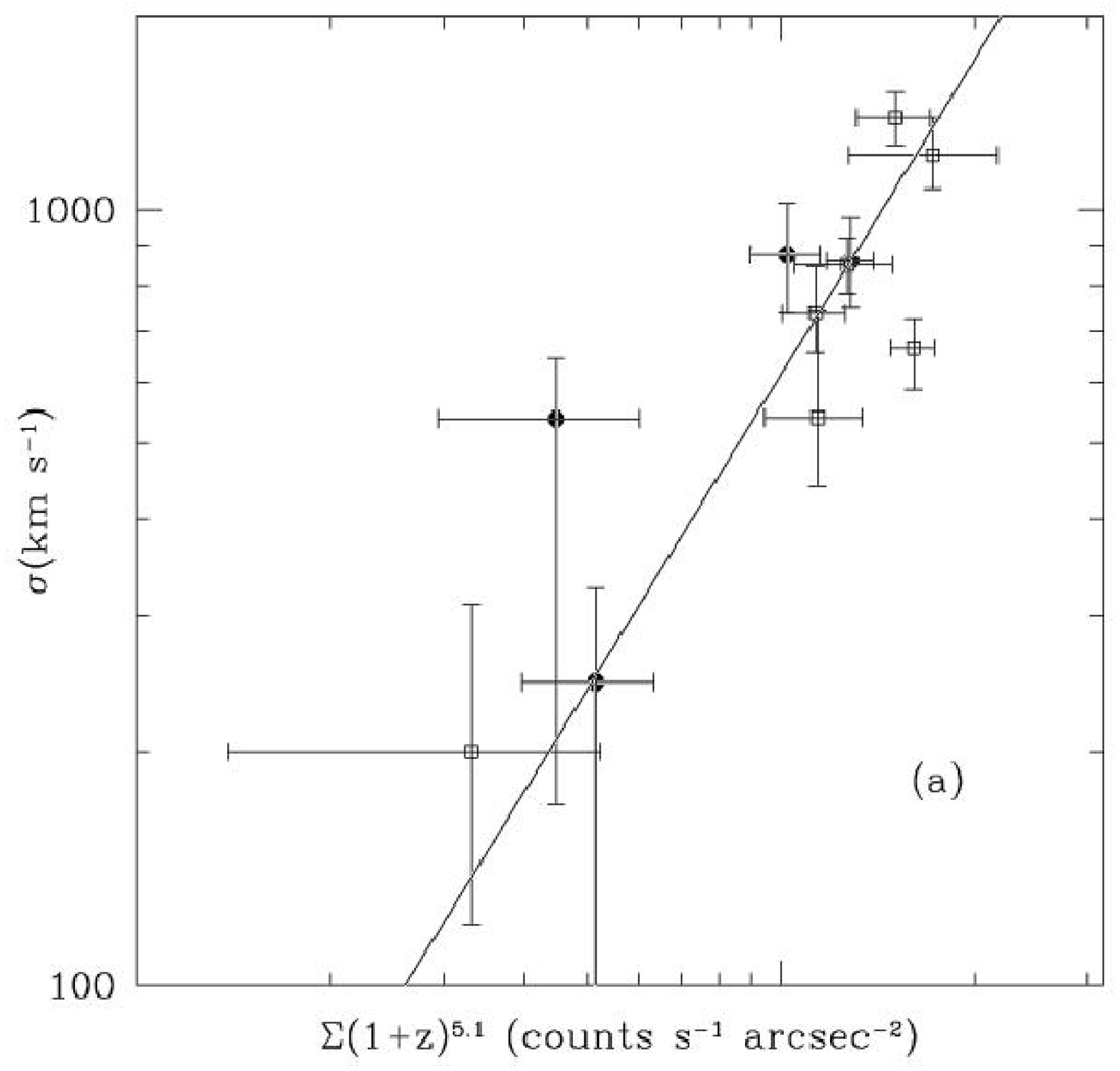}
\caption{X-ray temperature and galaxy velocity dispersion vs. 
optical central surface brightness, $\Sigma$}
\end{figure}

Selecting the BCG is often non-trivial because of contamination and
the definition of a selection radius. Due to space limitations we
do not discuss this issue further here (a full discussion 
is presented by Nelson {\it et al.} 2001a).

\section{Luminosity and Color Evolution and the Nature of the Accretion}

We show the I-band BCG Hubble diagram in Fig 2. In the lower panel
we bin the data along redshift to reveal the mean
trends. At $z <0.6$ the average BCG appears to follow the no-evolution
prediction, while at high redshifts they appear to begin to edge
toward the passive evolution model. We find that this peculiar
behavior is due to the inclusion of BCGs from a wide range of 
clusters. In particular, as we discuss next, we find that the BCGs
in clusters with $L_X > 2\times 10^{44}$ ergs s$^{-1}$ lie along the passive
evolution models, while the BCGs in fainter clusters lie along the
no-evolution models. The different relative fraction of the two types
of BCGs at different redshifts leads to the initially perplexing
Hubble diagram.

\begin{figure}
\plotfiddle{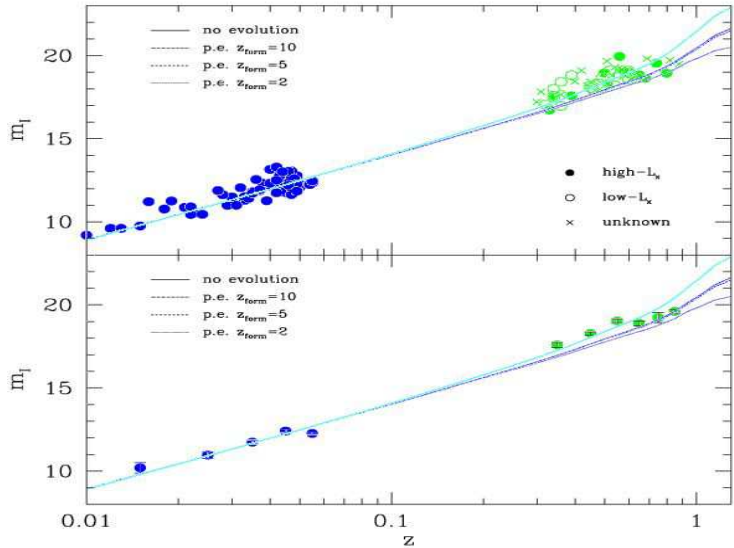}{185 truept}{0} {50}{38}{-150}{-70}
\caption{$I$-band BCG Hubble Diagram. Symbols in the upper panel 
represent individual BCGs, symbols in the lower panel represent
the means in redshift bins. The BCGs at $z < 0.1$ are from
Graham {\it et al.} Clusters with `unknown' values of L$_X$ at
$z > 0.6$ are almost certainly high-L$_X$ clusters because those
are the only ones sufficiently bright to be detected. The model
lines are described in the upper left of each panel (p.e. = passive
evolution).}
\end{figure}

Following the suggestion by Burke, Collins, \& Mann (BCM)
that $2 \times 10^{44}$ ergs s$^{-1}$ may
be an appropriate division of cluster types at these redshifts, we have divided our sample
at the corresponding value of the central surface brightness. 
Because at $z > 0.6$ our clusters fall primarily in the high $L_X$
sample, the high $L_X$ sample is consistent with passive evolution
models. The low L$_X$ sample, which is primarily our lower redshift
sample, is consistent with no-evolution models. 

Aragon-Salamanca, Baugh and Kauffmann (1998) interpreted consistency with 
no-evolution models to indicate the presence of accretion (because
no-evolution is non-physical). They inferred that these galaxies 
accreted 2 to 4 times their mass since a redshift of one. BCM found
that these results were derived primarily from a sample of BCGs
in low-L$_X$ clusters and concluded that much lower (by a factor of 2)
rates were upper limits on the accretion for BCGs in the more
massive environments. Qualitatively, we confirm both results
(accretion is necessary in BCGs in low-L$_X$ environments and
less  (or no) accretion is necessary in BCGs in high-L$_X$
environments).

\begin{figure}
\plotfiddle{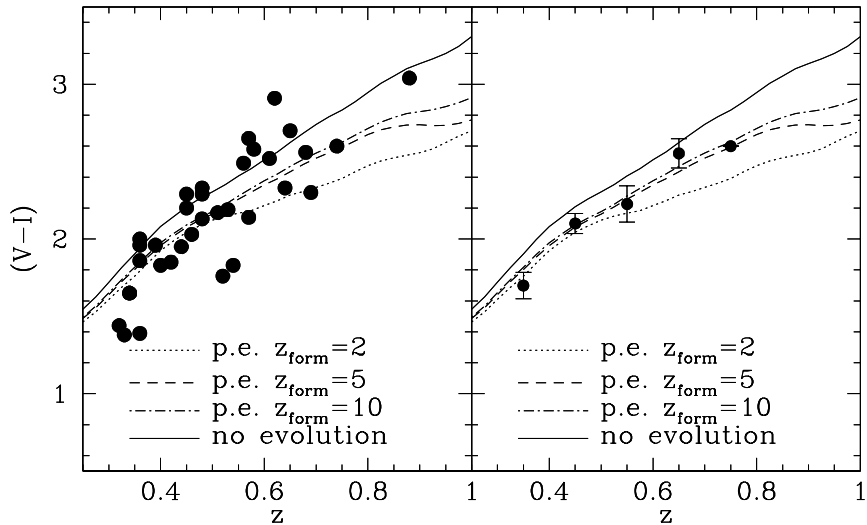}{220 truept}{-90}{50}{50}{-200}{270}
\caption{$V-I$ Color Evolution of BCGs. Left panel contains all of our
BCGs for which $V$ and $I$ data are available and the right panel 
contains the same data binned by redshift. The sole galaxies beyond
$z \sim 0.85$ is not included in the binned version.}
\end{figure}

Our data in $V, I$, and $K^\prime$ allow us to investigate the
color evolution of BCGs. In contrast to the situation for luminosity
evolution, the color evolution is consistent with passive evolution
for both the low and high L$_X$ samples (Figure 3). In all cases the stellar
populations of these galaxies appear to be rather
old, passively evolving. Therefore, whatever accretion is 
occurring, it consists of old stellar populations, not the accretion
of gas (i.e. a cooling flow) or of gas-rich galaxies, which would
presumably lead to a starburst. Interestingly, this result
fits in well with the recent discovery of red mergers at high
redshift (van Dokkum {\it et al.} 2000 in MS1054-03).

\section{The Search for Direct Evidence of Accretion}

\begin{figure}
\plotone{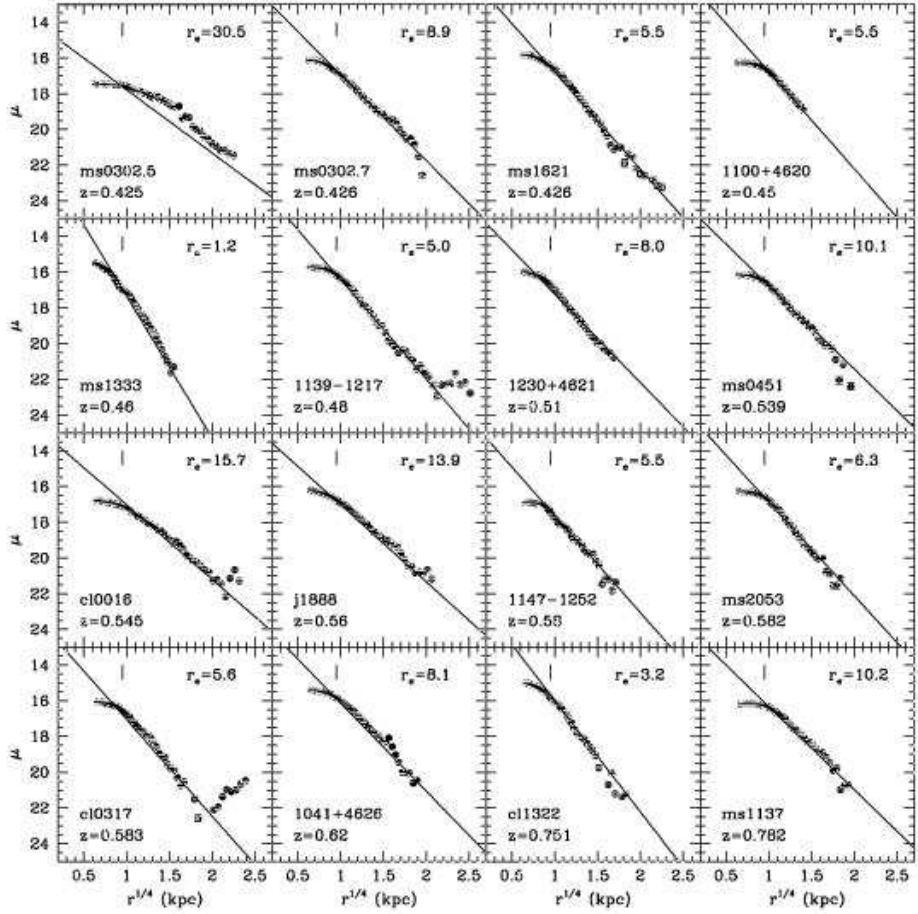}
\caption{H-band NICMOS Surface Brightness Profiles of High-z BCGs. The vertical
tick mark in each panel represents the resolution limit. }
\end{figure}

If significant accretion is occurring, 
can we see the structural properties of BCGs change with redshift?
If BCGs have velocity dispersions that are nearly independent of mass
(as seen locally Malumuth \& Kirshner 1981, 1985; Oegerle \&
Hoessel 1991), then changes in their masses must be
reflected by changes in their sizes. If BCG surface brightness
profiles can be characterized by deVaucouleurs profiles (see Graham {\it et al.}
1996; Gonzalez {\it et al.} 2000 for a variety), 
then a measure of their effective radii will provide 
a measure of their size. The nature of the profiles at large
radius is still a subject of debate (cf. 
Schombert 1986; Graham {\it et al.} 1996; Gonzalez {\it et al.} 2000), so 
it is possible that 
BCGs could change size and yet retain a constant scale length
at smaller radii (therefore, a null result with regard to scale
length changes as a function of redshift may not rule out accretion). 

\begin{figure}
\plotfiddle{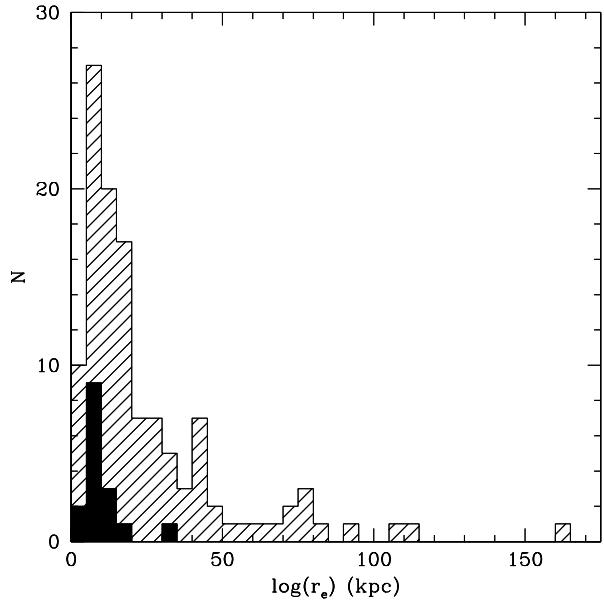}{200 truept}{0}{40}{40}{-120}{-70}
\caption{Comparison of Scale Lengths of Local and Distant BCGs. Hashed
histogram is the distribution of effective radii for local BCGs and the
filled histogram is the result from our NICMOS snapshots (Nelson {\it 
et al.} 2001).}
\end{figure}

Using NICMOS snapshots of 16 BCGs in a heterogeneous sample
of literature clusters at a mean redshift of $\sim 0.5$, 
to which we have fitted 
de Vaucouleurs profiles using the Gim2D software (Simard 1998), 
we measure the distribution of scale
lengths and compare to that measured from a local sample by
Graham {\it et al.} For all but one of the galaxies the r$^{1/4}$-law
profiles fit well (Figure 4). 
Comparison to similar measurements on archival
WFPC2 images of the BCGs for which some images are available, suggests
that non-uniform sky variations in the IR may lead us to 
underestimate the scale length by as much as 30\% systematically.
Although the high-redshift sample has a mean scale length that
is smaller than that of the local sample by somewhere between 20 and 50\%
(Figure 5), this result is preliminary because of the 
nagging systematic problems in resolving differences in scale lengths obtained from 
NICMOS and WFPC2 data (possibly due in part to physical effects
like color gradients). Neither the WFPC2 nor NICMOS images match the rest wavelength
of the local sample, and the differences we find between WFPC2 and NICMOS suggest
that scale length depends sensitively on color. The magnitude of the accretion
necessary to account for the luminosity evolution is within reach
of size measurements of BCGs, once the color-dependence of the scale-lengths is
understood.

\section{Conclusions}

Using a completely independent sample of high-redshift BCGs we have
confirmed the results of previous investigators: 1) BCG properties
vary as a function of environment, 2) BCGs in low-L$_X$ clusters
require significant accretion (a factor of 2 or more since $z \sim 1$)
to produce the observed Hubble diagram,
and 3) BCGs in high-L$_X$ clusters require much less (or possibly no)
accretion to produce the observed Hubble diagram. Additionally,
we find that the accretion must consist of old stellar populations
(because the colors of both BCG populations are consistent with
passive evolution models) and that the sizes of high-redshift
BCGs appear to be smaller than that of the local counterparts by
a factor that is consistent with the accretion model. Further study
of BCGs may provide the first truly direct observations of the
hierarchical growth of a galaxy population.

\acknowledgments

DZ acknowledges financial support from
the David and Lucile Packard Foundation, 
the Sloan Foundation, and the NSF CAREER program (AST 97-33111).



\begin{references}

\reference Aragon-Salamanca, A., Baugh, C.M., \& Kauffmann, G. 1998,
MNRAS, 297, 427
\reference Aragon-Salamanca, A., Ellis, R.S., Couch, W.J., \& Carter, D.
1993, MNRAS, 262, 764
\reference Burke, D.J., Collins, C.A., \& Mann, R.G. 2000, ApJ, 532L, 105
\reference Dalcanton, J.J., 1996, ApJ, 466, 92
\reference Gonzalez, A.H., Zabludoff, A.I., Zaritsky, D., \&
Dalcanton, J.J. 2000, ApJ, 536, 561
\reference Gonzalez, A.H., Zaritsky, D., Dalcanton, J.J., \& Nelson, A.E.
2001, ApJS, in press
\reference Graham, A., Lauer, T.R., Colless, M., \& Postman, M. 1996, ApJ,
465, 534
\reference Hoessel, J.G., 1980, ApJ, 241, 493
\reference Malumuth, E.M., \& Kirshner, R.P. 1981, ApJ, 251, 508
\reference Malumuth, E.M., \& Kirshner, R.P. 1985, ApJ, 291, 8
\reference Nelson, A.E., Gonzalez, A.H., Zaritsky, D., \& Dalcanton, J.J., 
2001a, ApJ., submitted
\reference Nelson, A.E., Gonzalez, A.H., Zaritsky, D., \& Dalcanton, J.J., 
2001b, ApJ., submitted
\reference Nelson, A.E., Simard, L., Zaritsky, D., Dalcanton, J.J., \&
Gonzalez, A.H. 2001c, ApJ., submitted
\reference Oegerle, W.R., \& Hoessel, J.G. 1991, ApJ, 375, 15
\reference Oemler, A. 1976, 209, 693
\reference Schombert, J.M. 1986, ApJS, 60, 603
\reference Simard, L. 1998, in
`Astronomical Data Analysis Software and Systems VII', ASP Conference Series,
145 (R. Albrecht, R. N. Hook and H. A. Bushouse, eds)
\reference Tremaine, S.D., \& Richstone, D. 1977, ApJ, 212, 311
\reference Zaritsky, D., Nelson, A.E., Dalcanton, J.J., \& Gonzalez, A.H.,
1997, ApJL, 480, L91
\reference Zaritsky, D., Shectman, S.A., \& Bredthauer, G. 1996, PASP, 108, 104
\end{references}
\end{document}